\documentclass[twocolumn,prl,preprintnumbers,amsmath,amssymb]{revtex4}

\newcommand{\be}
{\begin{eqnarray}}

\newcommand{\ee}
{\end{eqnarray}}

\usepackage{graphicx}
\usepackage{dcolumn}
\usepackage{bm}
\usepackage{times}

\begin{document}

\bibliographystyle{unsrt}

\title{Photon Antibunching in the Photoluminescence Spectra of a Single Carbon Nanotube}
\author{Alexander H\"{o}gele}
\author{Christophe Galland}
\author{Martin Winger}
\author{Atac Imamo\u{g}lu}
\affiliation{Institute of Quantum Electronics, ETH H\"{o}nggerberg, Wolfgang-Pauli-Strasse 16, CH-8093 Z\"{u}rich, Switzerland}

\date{published 27 May 2008}

\begin{abstract}
We report the first observation of photon antibunching in
the photoluminescence from single carbon nanotubes. The emergence of a
fast luminescence decay component under strong optical excitation
indicates that Auger processes are partially responsible for
inhibiting two-photon generation. Additionally, the presence of
exciton localization at low-temperatures ensures that nanotubes emit
photons predominantly one-by-one. The fact that multi-photon
emission probability can be smaller than 5\% suggests that carbon
nanotubes could be used as a source of single photons for
applications in quantum cryptography.
\end{abstract}

\maketitle

Carbon nanotubes provide a unique  paradigm  for mesoscopic physics
\cite{general}: while strong one-dimensional (1D) confinement
enhanced Coulomb interactions lead to striking new phenomena such as
spin-charge separation, absence of hyperfine interactions promise
ultra-long electron spin coherence for applications in quantum
information processing. Since the first observation of
photoluminescence (PL) from semiconducting single-walled carbon
nanotubes (CNTs) \cite{Connell}, a basic understanding of their
linear optical properties such as the role of strong exciton binding
\cite{Chang,Spataru,Perebeinos1,Zhao,Wang1,Maultzsch} and chirality
dictated optical excitations \cite{Bachilo1} have emerged. However,
all optical experiments on CNTs to date can be explained using
classical Maxwell's equations. Here, we report the first observation
of quantum correlations in PL from a single CNT.

It is well known that physical systems with quantum confinement of carriers in all directions such as atoms, ions, molecules, solid-sate quantum dots (QDs) or nitrogen vacancy centers in diamond exhibit photon antibunching \cite{Orrit}. The optical anharmonicity in such quasi zero-dimensional systems arises from phase-space filling - a consequence of the Pauli exclusion principle which prevents carriers from occupying identical quantum states. It is a priori not evident how the softening of the confinement potential in one direction will affect the statistics of photon emission events. Solid-state systems extended in one dimension allow for the coexistence of excited electron-hole pairs which in turn by radiative recombination generate multiple photons at a time. On the other hand, nonlinear exciton-exciton annihilation enhanced by inefficient screening of the Coulomb interactions in 1D could be effective enough \cite{Ma,Wang-Auger} to ensure quantum light emission governed by sub-Poissonian statistics. In particular, Auger processes could enforce non-radiative recombination upon excitation of a second exciton, thereby inhibiting simultaneous two-photon emission events.

In the present work we experimentally investigated the statistics of low-temperature PL emitted by single CNTs. We have studied single-wall CoMoCat nanotubes encapsulated in sodium dodecylbenzenesulfonate which were deposited on a substrate out of an aqueous suspension. Silicon dioxide substrate was used for sample characterization with atomic force microscopy (AFM). For optical studies, the CNTs were deposited directly onto the flat surface of a solid-immersion lens with a density below one nanotube per $\mu$m$^2$ and studied with an optical confocal microscope system in a liquid nitrogen dewar or a helium bath cryostat at 77~K and 4.2~K base temperatures, respectively. Temperature fine-tuning was achieved locally through a combination of a resistive heater and sensor. We focused our studies on individual CNTs with PL emission in the spectral window between 855~nm and 885~nm. It is most likely that in this spectral window the observed low-temperature PL originates from (6,4) chirality CNTs which have an emission wavelength of 873~nm at room temperature \cite{Bachilo1}. The laser excitation of a phonon sideband \cite{Jiang,Perebeinos-G,Chou,Htoon-PLE} 70~meV above the peak emission energy was carried out using a Ti:Sapphire laser in continuous wave operation for PL, and fs--pulsed mode (130~fs pulse duration, 76.34~MHz repetition rate) for both time-resolved and photon correlation measurements. Our experimental PL setup had a spectral resolution of 0.3~meV, the temporal resolution of the avalanche photodiodes used in the Hanbury-Brown and Twiss setup was 300~ps. A streak-camera with a temporal resolution of 7~ps was used for time-resolved PL spectroscopy.

\begin{figure}[t]
\includegraphics[scale=0.8]{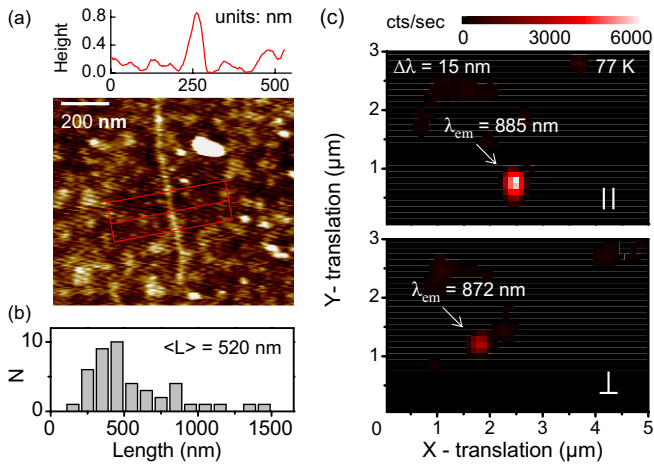}
\caption{Characterization of CoMoCat carbon nanotube samples by AFM (a,b)
and optical confocal microscopy (c). The topography image shows an isolated nanotube with a height profile of $0.8$~nm (measured in the region indicated by the rectangle). The length of the nanotube is $800$~nm. The length distribution (b) of nanotubes was obtained from the analysis of different samples with AFM. The mean nanotube length in our samples is $<L>=520$~nm. Local photoluminescence map (c) of an area of $3\times5~\mu$m$^2$ recorded for two orthogonal linear polarizations (indicated by symbols) of the excitation laser. The false color image was acquired in a spectral bandwidth of 15~nm at 77~K by scanning the sample in discrete steps of 100~nm relative to the optical spot of 450~nm diameter. The two bright emission spots in the upper and lower graph originate from individual nanotubes with an emission wavelength of 885~nm and 872~nm, respectively.}
\end{figure}

\begin{figure}[t]
\includegraphics[scale=0.8]{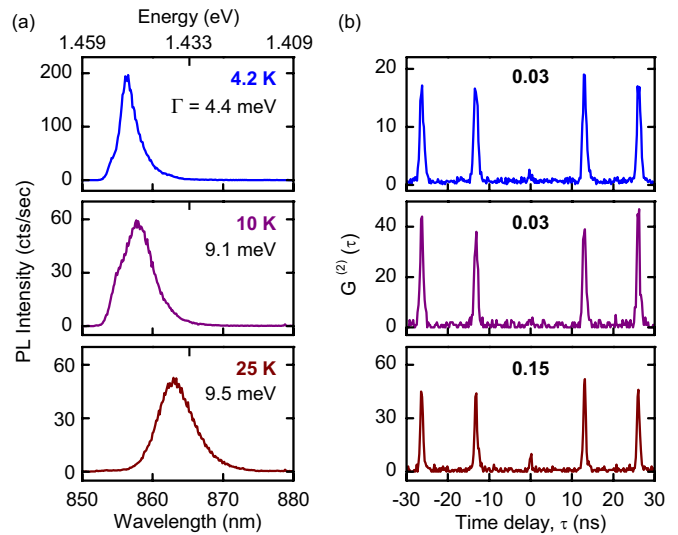}
\caption{Photoluminescence and photon correlation spectra of a single carbon nanotube at three different temperatures. The nanotube was excited with a Ti:sapphire laser through a phonon sideband $70$~meV above the emission energy in cw mode and fs--pulsed mode for PL and photon correlation measurements, respectively. The left panel (a) shows the photoluminescence intensity as a function of wavelength for a given temperature. The right panel (b) depicts for each temperature the corresponding unnormalized correlation function $G^{(2)}(\tau)$ measured in a Hanbury-Brown and Twiss setup. The strongly inhibited correlation signal of the central peak at $\tau=0$ indicates the suppressed two (or more) photon emission probability (given as numbers of the normalized correlation function $g^{(2)}(0)$ in the right panel) and is a hallmark of nonclassical light.}
\end{figure}

Fig.~1a shows an AFM topography scan of a $\sim\mu$m$^2$ area with a single nanotube on silicon dioxide. It has a measured height of 0.8~nm in agreement with the average CNT diameter of 0.81~nm characteristic for CoMoCat nanotubes \cite{Bachilo2}. AFM analysis of several samples yielded the length distribution of CNTs (Fig.~1b) with an average length of $<L>=520$~nm (the median was 420~nm; short nanotubes were underrepresented since they were hardly distinguishable from unintentional nanoscale deposits). The false color images in Fig.~1c represent local low-temperature PL emission from CNTs deposited on the solid-immersion lens. The luminescence was integrated in a narrow wavelength band of 15~nm around 880~nm, the map of $3\times5~\mu$m was acquired for two orthogonal orientations of the linear excitation polarization (upper and lower graph in Fig.~1c). Two bright spots at different locations of the sample represent emission from two individual nanotubes at 885~nm and 872~nm with orthogonal linear excitation axes. The nanotubes showed strong selectivity to linear excitation polarization - the well-known antenna effect characteristic of 1D emitters. The nanotube emission was spatially localized with a characteristic length scale below our spatial resolution (focal spot full-width at half-maximum $\sim 450$~nm), a typical feature of our sample. Both the localized emission and the strong selectivity to linearly polarized excitation are consistent with previous single CNT PL studies \cite{Hartschuh}.

Fig.~2a presents typical PL spectra from an individual CNT. With decreasing temperature the photoluminescence resonance exhibited a blueshift and evolved towards an asymmetric line with a steep high-energy shoulder, while the full-width at half-maximum linewidth $\Gamma$ decreased down to 4.4~meV at 4.2~K. Asymmetric and multi-line PL spectra were found in previous low-temperature single CNT studies \cite{Htoon-4K,Lefebvre1,Hagen}. Right panel of Fig.~2 shows the results of photon correlation measurements carried out on the same CNT using pulsed-laser excitation and directing the PL to a Hanbury-Brown and Twiss setup without spectral filtering: the missing central peak in the unnormalized correlation function $G^{(2)}(\tau)$ at $\tau = 0$ is a direct signature of strong photon antibunching, indicating that this CNT essentially never emits two photons upon excitation by a fs laser pulse \cite{Michler-Science}. Similar measurements carried out on an attenuated laser naturally yield a $\tau = 0$ peak that is as strong as all the other peaks; reduction of the normalized area of the central peak below unity provides clear evidence that the CNT PL is nonclassical. The degree of antibunching is substantial, two-photon emission probability remains as low as 3\% up to 10~K. Even at 25~K the CNT emits two photons at a time with a probability of only 15\%. All of the CNTs that we investigated at low temperatures exhibited photon antibunching, however with normalized correlation peaks $g^{(2)}(0)$ varying between 3\% and 40\% at 4.2~K. The finite value of the peak at $\tau=0$ as well as its variations were partially related to the level of background at different sample locations but also due to excitation powers set close to saturation for weakly emitting CNTs. A general feature was that nanotubes with a single PL resonance showed stronger photon antibunching at low excitation powers.

\begin{figure}[t]
\includegraphics[scale=0.8]{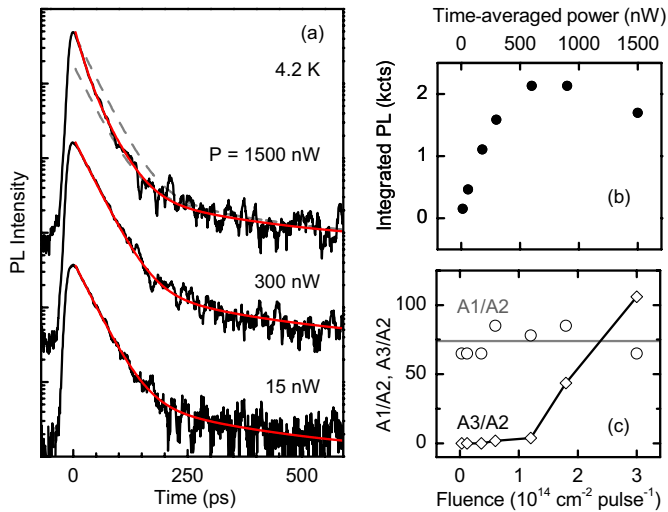}
\caption{Low-temperature  time-resolved photoluminescence data
recorded for pulsed excitation of a single CNT with PL peak emission
at 860~nm. The PL intensity saturates with increasing exciatation
power, as plotted in (b). At low excitation powers two decay
processes with characteristic decay times of 35~ps and 250~ps are
present. The average ratio of the amplitudes A1 and A2 associated
with the strength of the 35~ps and 250~ps decay processes,
respectively, is 74 [grey solid line in (c)] and independent of pump
power [open circles in (c)]. A third, fast decay component of 15~ps,
emerges at high pump powers; its amplitude A3 is shown as open
diamonds in (c) normalized to the slowest decay amplitude A2 (the
black solid line is a guide to the eye). The corresponding bi- and
tri-exponential decay fits (red solid lines) to the TRPL intensity
(data offset for clarity) are shown in a logarithmic plot (a), grey
dashed lines represent best bi-exponential trial fits to the high
power spectrum omitting the fast 15 ps decay component.}
\end{figure}

\begin{figure}[t]
\includegraphics[scale=0.8]{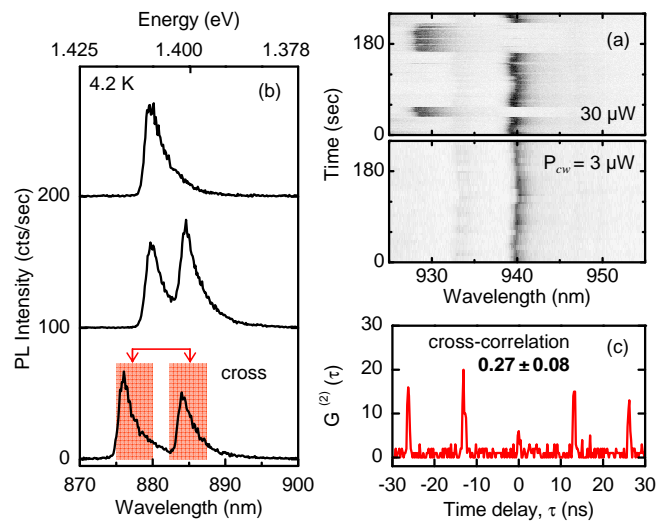}
\caption{Time evolution of single carbon nanotube photoluminescence at 4.2~K for excitation powers (continuous-wave) of $30~\mu$W (a, upper panel) and $3~\mu$W (a, lower panel) are shown in grey-scale representation (1~sec integration time per spectrum). High excitation power causes spectral jitter of the luminescence between two states separated by $\sim 10$~meV on a timescale of minutes. The jitter was absent for low excitation powers. Photoluminescence spectra (b, offset for clarity) and photon cross-correlation spectrum (c) of another single carbon nanotube. Starting with a single resonance (upper panel) the nanotube emission spectrum exhibited a splitting of 7~meV after one day of observation (central panel) and a further spectral shift of 6~meV and 1~mev for the higher and lower energy peak, respectively, after the second day (lower panel). The two emission peaks were spectrally selected with band-pass filters (10~nm bandwidth) for cross-correlation (c) and auto-correlation measurements (not shown).}
\end{figure}

In order to examine the excitation power dependence and the potential role of Auger mediated exciton-exciton annihilation, we investigated a CNT with a single PL line by time-resolved PL spectroscopy. At low excitation powers we observed two characteristic PL decay times. We find for this nanotube, as for most of the CNTs at 4.2~K, a dominant low-excitation-power decay time constant ranging between 20~ps and 40~ps (Fig.~3a, upper panel), in agreement with previous reports \cite{Hagen,Hirori}. These observations are also in good agreement with what one would
expect from theoretical estimates ($10-20$~ps) of radiative lifetime of non-localized CNT excitons \cite{Spataru-lt}. We further observe a two orders of magnitude weaker contribution to time-resolved PL signal from a slow decay process (250~ps time constant). Increasing the pump power leads to saturation of the CNT exciton emission (Fig.~3b); this is a signature characteristic of quantum emitters, such as atoms or QDs \cite{Brunner}. On the other hand, for high pump intensities, we observe the emergence of a third, fast decay component of 15~ps (Fig.~3a, lower panel) which eventually dominates the PL decay (Fig.~3c). This nonlinear decay process is not typical for atoms or molecules but has signatures of an Auger-mediated exciton-exciton annihilation process. The measured Auger rate is an order of magnitude smaller than what was inferred from pump-probe experiments on CNTs of comparable length at room temperature \cite{Wang-Auger}. It is likely that temperature driven exciton localization is responsible for the slowing down of the Auger rate, an effect of reduced dimensionality that was observed for colloidal nanoparticles \cite{Htoon-Auger}. We note that localization should give rise to two competing effects: on one hand, exciton confinement is expected to increase the local wavefunction overlap and thus the intra-site Auger rate. On the other hand, the overlap integral at different sites along the nanotube axis should be suppressed and the strength of inter-site nonlinear interactions would be reduced. It is obvious that the intra-site nonlinear Auger processes would play an important role in inhibiting the coexistence of two excitons and ensuring that the CNT behaves as an anharmonic quantum emitter \cite{Michler-Nature}. However, the observed strong antibunching cannot be explained by Auger processes alone since the corresponding rate is only twice the characteristic PL decay rate: for the particular case of the nanotube in Fig. 3, the antibunching decreased only slightly from $g^{(2)}(0)=0.10$ at low pump power to $g^{(2)}(0)=0.13$ in saturation. This is
yet another indication that exciton localization at low-temperatures is important for nonclassical CNT emission.

Fig.~4 presents a set of data that further supports the picture of exciton localization in CNTs. The formation of localized excitonic \cite{Hagen,Hirori} or QD-like states \cite{McEuen} at low temperatures is neither intentional nor controlled in our sample but characteristic for 1D systems which are sensitive to disorder. We observed variations in PL of single nanotubes both in terms of intensity and spectral position, just like in spherical \cite{Nirmal} and elongated \cite{Muller} colloidal nanocrystals. Fig.~4a exemplifies this behavior in a grey-scale plot: it shows a jitter on a time scale of minutes of both PL intensity and wavelength which become prominent as the laser intensity was increased. The fluctuations were also found to be sensitive to temperature cycling. Some CNTs that were exposed to intense laser excitation, as was typically the case in pulsed photon correlation experiments, showed a drastic change of PL characteristics. An example of an irreversible splitting of the initially single PL line is shown for an individual nanotube in Fig.~4b. After a day of laser exposure we observed the emergence of a satellite peak with comparable intensity (Fig.~4b, central panel), and an increased energy separation of the two peaks after another day of observation (Fig.~4b, lower panel). For this nanotube, however, we did not observe distinct spectral jumps (on time scales down to a second) in continuous-wave PL experiments, in contrast to the CNT in Fig.~4a. We interpret multi-resonance PL and its jitter in time in terms of exciton localization sites created along the nanotube axis with energies determined by the quantum confined Stark effect in an unstable electrostatic environment, features well known from photophysics of single colloidal QDs \cite{Empedocles}. In our experiments, rearrangement of the electrostatic environment is attributed to high laser excitation intensity, the details of underlying mechanisms, however, are unknown and require more systematic studies.

We have employed photon correlation spectroscopy \cite{Kiraz02} to study the photon statistics of the CNT depicted in Fig.~4b. Under pulsed excitation, the photon auto-correlation of the combined two-peak PL gave an antibunching of $0.42 \pm 0.07$ suggesting two uncorrelated quantum emitters. This is ruled out, however, by auto-correlation measurements of the lower and higher energy peaks which showed only moderate antibunching of $0.39 \pm 0.08$ and $0.26 \pm 0.06$, respectively. Remarkably, photon cross-correlation function of the two PL lines, presented in Fig.~4c, showed an antibunching of $0.27 \pm 0.08$. This observation proves that the two emission lines are quantum correlated, which in turn can be explained with two different scenarios. One explanation is based on two quantum correlated yet spatially separated localized sites. The observation that the relative intensity of the two peaks was dependent on the lateral displacement of the sample with respect to the laser spot supports this picture. A second explanation relies on one QD-like emission site having two temporal states of different energies due to bistable neighboring charge configuration, as in Fig.~4a. For the latter scenario the non-zero central peak in cross-correlation implies that the switching between the two distinct temporal states must occur on a time scale shorter than the characteristic PL decay time of $20-40$~ps. In the limit of two uncorrelated quantum emitters or one quantum emitter but slow switching events, the cross-correlation measurement should yield a vanishing central peak.

Our experiments demonstrate for the first time that a CNT emits nonclassical light at low temperatures. The results of time-resolved PL and photon cross-correlation measurements establish the role of exciton localization and Auger decay of multiexciton states leading to quantum correlations between photon emission events. An intriguing question is whether exciton-exciton interactions could give rise to quantum correlations in delocalized 1D or spatially separated QD exciton systems. With respect to applications, our findings could motivate the development of new single-photon sources for long-distance quantum communication since luminescence in CNTs can be generated electrically \cite{Avouris} and larger diameter nanotube emission extends into the optical communication window.

We acknowledge R. McKenzie and B. Babic for help with sample preparation as well as C. Latta and P. Maletinsky for support and collaboration in the lab. A. H. acknowledges K. Karrai for helpful discussions. This work was supported by a grant from the Swiss National Science Foundation (SNSF).

\end{document}